\documentstyle[11pt,epsfig]{article}
\newcommand{\be}{\begin{equation}}
\newcommand{\ee}{\end{equation}}
\newcommand{\bi}{\bibitem}

\begin{document}
{\hbox to\hsize{August, 1998 \hfill TAC-1998-018}}
\vglue .06in
\begin{center}{
{\Large \bf { Superluminal propagation of light in gravitational field 
and  non-causal signals.  }}
\bigskip
\\
{\bf A.D. Dolgov 
\footnote{Also: ITEP, Bol. Cheremushkinskaya 25, Moscow 113259, Russia.}
and I.D. Novikov
\footnote{Also: University Observatory, Juliane Maries Vej 30, DK-2100, 
Copenhagen {\O}, Denmark; \newline
Astro Space Center of Lebedev Physical Int,  
Profsoyuznaya 84/32, Moscow 117810, Russia; \newline
NORDITA, Blegdamsvej 17, DK-2110, Copenhagen {\O}, Denmark.}
 \\[.05in]
{\it Teoretisk Astrofysik Center\\
 Juliane Maries Vej 30, DK-2100, Copenhagen {\O}, Denmark
}}\\
}\end{center}

\begin{abstract}
\setlength{\baselineskip}{0.3in}
It has been found in several papers that, because of quantum corrections,
light front can propagate with superluminal velocity in gravitational fields 
and even in flat space-time across two conducting plates. We show that, if 
this is the case, closed time-like trajectories would be possible and, in 
particular, in certain reference frames photons could return to their source 
of origin before they were produced there, in contrast to the opposite claim 
made in the literature. 

\end{abstract}
\bigskip
\setlength{\baselineskip}{0.3in}

\section{Introduction}  

A study of photon propagation in gravitational field has revealed a surprising 
phenomenon that quantum corrections modify the characteristics of photon 
equation of motion in such a way that in some cases they may lay outside
the light cone. This effect was first found in ref. \cite{dh} for several
different geometries (gravitational wave, Schwarzschild, and Robertson-Walker).
In subsequent papers it has been shown that photons may propagate "faster 
than light" also in the Reissner-Nordstr{\H o}m \cite{ds1} 
and the Kerr \cite{ds2} backgrounds. Similar effect was found for  
propagation of massless neutrinos in gravitational field \cite{yo}. The only
essential difference is that photons can propagate with $v>c$ even in vacuum
where Riemann tensor is non-vanishing, $R_{\alpha\beta\mu\nu} \neq 0$, while
neutrinos  may acquire superluminal velocity only in space-time with 
$ R_{\mu\nu} \neq 0$. Superluminal propagation has been also found for
flat space-time with boundaries, e.g. for photon propagation between
conducting plates \cite{ks,gb,lpt} (see also \cite{sbm}). In what follows we 
will discuss both possibilities and will show that in these cases one could
find a coordinate frame where photons would return to their source
before they were created there. 

The photon effective action in vacuum in one loop approximation and in the 
lowest order in the gravitational field strength is described by 
the well known vacuum
polarization diagram. Because of gauge invariance of electrodynamics and 
general covariance of gravity the result for the action is unique and
can be written immediately:
\be 
S = \int d^4 x \sqrt {-g} \left( -{1\over 4} F_{\mu\nu}F^{\mu\nu}
+ {\alpha C \over m_e^2} R_{\alpha\beta\mu\nu}F_{\mu\nu}F^{\alpha\beta}
\right)
\label{s}
\ee
Here $F_{\mu\nu}=\partial_\mu A_\nu - \partial_\nu A_\mu$ is the 
electromagnetic field tensor,
$\alpha=1/137$ is the fine structure constant, $m_e$ is the mass of
electron and the coefficient $C$, as calculated in ref. \cite{dh}, is 
$C= - 1/360\pi$. For our purpose is essential just that $C\neq 0$, even its 
sign is not important. The action may also contain other terms proportional to 
the Ricci tensor $R_{\mu\nu}$ or to the curvature scalar $R$ but they both
vanish in vacuum and will be neglected in what follows. 

It is evident that this action leads to the equation of motion with modified
highest derivative terms and generally speaking the coefficient at
$\partial^2 /\partial_t^2$ would not be equal to the one at
$\partial^2 /\partial_j^2$ (with the evident factor $g_{ij}/g_{tt}$). 
In other words the characteristics of the photon wave equation with the
corrections (\ref{s}) generically would not coincide with the normal 
light cone. It means that the velocity of the propagation of the {\it front}
of the signal would be modified.
One should not worry if this happened to be inside the light cone
but as we have already mentioned this is not the case. In particular, as was
shown in ref. \cite {dh}, in the Schwarzschild background photons going
in a non-radial direction could propagate either faster than $c$ or slower
depending on their polarization. This may imply serious problems for the 
theory.

The velocity of light front propagation is known \cite{lm}
to be determined by the asymptotics of the refraction index $n(\omega)$ for
$\omega \rightarrow \infty$. Since the action (\ref{s}) is believed to be
valid only for sufficiently low photon frequency, $\omega \ll m_e$, the 
problem of superluminal propagation might be resolved by possible 
unaccounted for terms which vanish in the low frequency limit. It was
argued in ref. \cite{dh} that if the refraction index satisfy
dispersion relation with a positive definite $Im \, n(\omega)$, then 
$n(0) > n(\infty )$ and the problem persists.  However it was shown \cite{dk}
that positiveness of $Im \, n $ in gravitational fields is not necessarily 
true and one may hope that high frequency contributions 
would give rise to $n(\infty ) > 1$ and correspondingly $v<c$. 

Another twist to 
the problem was given by observation \cite{ibk} that the expression for 
refraction index calculated from the effective action
(\ref{s}) is valid in the first order of gravitational interaction for
{\it any} value of photon frequency. Indeed refraction index is determined by
the forward scattering amplitude and the latter, in turn, is exactly given by
the second term because forward scattering amplitude of a photon in the lowest 
order in external field can be only a function of photon 4-momenta squared,
$k_1^2 =k_2^2 =0$ and of the transferred momentum, $q^2 = (k_1 -k_2)^2$, which
also vanishes for forward scattering. This simple argument invalidates the
possibility of resolving the problem by higher order terms in $\omega/m_e$
mentioned in references \cite{dk,gms}. 

Thus in the lowest order in external classical
gravitational field, in the first order of electromagnetic coupling $\alpha$,
and neglecting quantum gravity corrections the result~(\ref{s}) determines
photon refraction index for all frequencies. In this approximation photons
in a certain polarization state would propagate outside the normal light cone.
In this connection the following two questions arise. First, if such 
superluminal propagation would create any problem with causality and, second,
if higher order corrections in electromagnetic and gravitational interactions
could return photons "back to normality". 

The first question has been addressed
in the literature (in practically all quoted here papers) and the almost
unanimous conclusion is that, though there exist reference frames in which 
superluminal photons would reach detector, by the clock of an observer in this 
frame, earlier than they were emitted but it is impossible to send photons 
back to their source prior to their emission in the proper time of the source.
In other words, non-causal signals would not be possible and, in particular, 
closed space-like trajectories for superluminal photons
do not exist. This conclusion was based
on the absence of Poincare invariance of the theory due to presence of a fixed
gravitating center. We will show here that this conclusion is incorrect
and explicitly construct an example of a non-causal 
closed space-like trajectory for a
signal propagating outside the light cone. In view of that the
problem seems to be considerably more grave than it was apprehended earlier. 

Before presenting our example let us discuss what mechanisms or what
changes in physics may in principle cancel the effect of superluminal
propagation or prevent from traveling backward in time if the effect 
persists. An evident possibility is a contribution from higher order 
corrections in electromagnetic and gravitational interactions. Still it is
not easy to achieve. Higher order electromagnetic corrections do not help.
They could only give an extra power of $\alpha$ and no extra terms in the 
action proportional to derivatives of $F_{\mu\nu}$. It follows from 
dimensional consideration and from renormalizability of quantum electrodynamics
in classical gravitational background. The structure of correction to the
effective action remains the same, $\delta S \sim (RFF)/m_e^2$. Power 
counting shows that the diagrams with several external graviton legs (higher
order corrections in external field) also do not give rise to derivative of
electromagnetic field. The only dangerous diagrams are those with virtual 
gravitons, and only with at least two virtual gravitons. Such diagrams could 
give the terms of the following form:
\be
\delta_2 S = C_2 (\partial F \partial F R)/ m^4_{Pl}
\label{ds2}
\ee
Here $m^4_{Pl}$ is the Planck mass and in the expression in the brackets 
a proper contraction of indices is made.

Since the coefficient $C_2$ cannot be large (in fact it should be much 
smaller than unity), this contribution is not dangerous too for photon
frequency below $m_{Pl}$. The loops with three 
virtual gravitons can give terms similar to (\ref{ds2}) but now with the 
diverging coefficient $C_3 \sim \Lambda^2 /m^4_{Pl}$. Strictly speaking one 
cannot say anything about the magnitude of such terms but in superstring based
theories of consistent renormalizable quantum gravity one would expect that 
$\Lambda^2 \sim m^2_{Pl}$ and in these theory (theories?) the contribution
of higher order corrections from virtual gravitons would be negligible. Still
it is worthwhile to understand  more rigorously if the effect of 
superluminal propagation exists in a well defined quantum gravity. 

Another possibility mentioned in the literature \cite{dh}  is a modification
of the usual causal light cone by an effective cone which corresponds to
the fastest possible signal propagation. This would  require quite serious 
changes in the usual physics and moreover it is not clear how it could be 
realized. The theory even with the modified effective action (\ref{s}) 
remains Lorenz invariant as well as general covariant. It permits to 
express time running in a new reference frame through time and coordinates in
the original frame. Introduction of a new  causal cone with a non-constant 
maximum speed would probably result in a theory quite different from 
General Relativity. So we have either to admit that general relativity is
broken at velocities very close to speed of light or to live in the world
with non-causal signals.

\section{Non-causal signals in Special Relativity}

Let us remind how one can obtain acausal signal propagation in Special
Relativity with "normal" tachyons moving with a constant speed $u$,
exceeding speed of light $c$ (which  we take to be equal to 1). We assume that
a tachyon is emitted at the point $x_1$ at the moment $t_1$ and is
registered at the point $x_2$ at the moment $t_2$ of some inertial frame. 
In this reference frame the evident relation holds:
\be
t_2 - t_1 =  (x_2 -x_1)/u >0
\label{t2t1}
\ee
Now let us make Lorenz transformation to another inertial frame moving
with respect to the first one with velocity $V$:
\be
x' = \gamma (x-Vt), \,\,\,\, t' =\gamma (t-Vx)
\label{x'}
\ee
where $\gamma = 1/\sqrt{1-V^2}$. 

In the new frame the time interval between emission and registration of the 
tachyon is given by:
\be
t'_2 - t'_1 = \gamma (t_2 - t_1 ) (1 - V u)
\label{t'2}
\ee
For $V= 1/u$ the time interval $(t'_2 - t'_1)$ can be zero, which
corresponds to tachyon propagating in the second frame with infinite velocity 
and for a larger $V$ the interval $(t'_2 - t'_1)$ could be even negative, i.e.
the tachyon in this frame propagates backward in time. The absolute value of 
the tachyon velocity is always bigger than $c$, approaching $\pm \infty$ when
$V$ tends to $1/u$ from above or from below. Let us assume now that 
at the moment when tachyon reaches the detector at the point $x_2$ (or $x'_2$
in the second frame) this detector emits another tachyon back to the source. 
Let us assume for simplicity that the picture is symmetric so that 
the second source/detector emits a tachyon with the same velocity $u$ 
with respect to itself and that both the
first and the second emitters of tachyons move in the primed reference frame 
in opposite directions with equal velocities $V=1/u$. Thus both tachyons 
would have infinite velocities in this frame and the signal would instantly 
return to the place of origin. It is evident that in the case of $V>1/u$ 
both tachyons would travel into the past and the signal would return back  
to the first emitters "in less than no time".

Bearing in mind the examples presented below we will consider a slightly
modified and more "realistic" construction of gedanken experiment with 
tachyons, which permits to avoid collision of tachyon emitters. Let us assume
that there are two identical tachyon emitters A and B moving in the opposite 
directions along parallel straight lines (along $x$) separated by some distance
$\Delta y$ (see fig. 1). 

A tachyon is emitted by the source A at the time moment $t_0$ and in the
chosen reference frame it moves backward in time. At the moment $t_1$
($t_1 <t_0$) the tachyon reflects in perpendicular direction (now it moves
along $y$) and at the moment $t_2$ ($t_1<t_2<t_0$)
it collides with the emitter B. Space-time
picture of its motion (in terms of $t$ and $x$ with fixed $y$) is presented in
fig. 2. The collision of the tachyon with B triggers the emission by the 
latter of another tachyon. This new tachyon moves again backward in time
along the line of motion of B but in the opposite direction (all along $x$).
At the moment $t_3$ it reflects perpendicular to $x$ and moves to A and
hits the latter at the moment $t_4$. It is evident that the system can be 
chosen so that $t_4< t_0$. Space-time picture of the motion of the second
tachyon is presented in fig. 3.

The possibility of sending signal, moving faster than light, 
back into the past to the source of its origin, which we have just demonstrated,
is well known in the standard Special Relativity. It is normally assumed
that tachyons move with a constant speed $u>1$ independently of space points.
In the case of propagation across two conduction plates it is not so,
the velocity of photon can be bigger then $c$ between the plates and equal to
$c$ outside \cite{ks,gb}. Still it is evident that the conclusion of acausal 
propagation would also survive in this case. 
Let us consider one dimensional motion along $x$ and assume that
the velocity of light is $v_l =1 $ for $|x| > d$ and $v_l =u >1$ for $|x|<d$.
For a realization of the gedanken experiment which we will discuss, a small
hole should be made in the plates so that the photon could penetrate into
the inner space. The size of the hole should be large in comparison with the
wave length and simultaneously small not to destroy the effect of superluminal
propagation. We will not go into these subtleties and consider this model as a
toy model for illustration of a possible travel into the past. The
arguments go essentially along the same lines as for "normal" tachyons.
Let us assume as above that photon is emitted at the point $x_1$ at the moment
$t_1$ and registered at $x_2$ at the moment $t_2$ in some inertial frame.
On the way the photon passes
the region between the plates where it moves faster than light with velocity
$u>1$. In this reference frame the following evident relation
holds:
\be
t_2 - t_1 = x_2 - x_1 - 2d + 2d/u
\label{t2plate}
\ee
We can again make the Lorenz transformation (\ref{x'}). The time of arrival 
of the signal in the primed coordinate frame to the point $x_2'$ is
given by the expression
\be
t'_2 -t_1' = \gamma (t_2 - t_1 ) \left( 1-V - {2d \over t_2 - t_1}\, 
{u-1 \over uV}\right)
\label{t2'plate}
\ee
Again the difference $(t'_2 -t_1')$ may be negative for $V$ sufficiently close
to unity. It is interesting that the sign of the ratio 
$(t'_2 -t_1')/(t_2 - t_1)$ depends upon the distance between the points 1 and 2.

The gedanken experiment with the return of the signal back to the emitter
prior to the emission can be constructed in the same way as above with 
the standard tachyons which moves with a constant superluminal velocity. 
The space-time picture for such an "experiment" is presented in fig. 4.
The world line of the signal in the primed
system looks as following: the photon is "produced" out of nothing somewhere
between his real birth place and the detector (the moment $t''$ in the
figure) and propagates both ways to the place of birth and to the detector. 
When it reaches the moment $t_0$ another photon (original) is emitted from the 
source and "annihilates" with the first one at the moment $t'$.  

Returning to arranging the time machine for photons in this conditions we 
can do essentially the same things as in the previously considered case with 
the only difference that now we have to supply both source/detectors A and B
with their own plates with holes so that one set of plates
moves together with A while the other moves with B . 

\section{Acausal signals in gravitational fields}
  
In the case of superluminal photons traveling in a gravitational field
the excess of the velocity over $c$ is not given by a step function but
continuously drops as $1/r^4$ with the increasing distance from the 
gravitating  center. Evidently
this does not create any serious difficulties. What is more important is the
curvature of space-time and the dependence of distance and time intervals
upon the space-time metric and also the bending of the tachyon trajectory and
the delay of signal due to interaction with gravitational field.

We will consider motion in spherically 
symmetric Schwarzschild background created by a localized matter not 
necessarily forming a black hole. The metric can be e.g. written in the form:
\be 
ds^2 = a^2 (r) dt^2 - b^2 (r) (dx^2 + dy^2 + dz^2)
\label{ds21}
\ee
where $r^2 = x^2 + y^2 + z^2$ and $a^2 = (1- r_g/4r)^2/(1 + r_g/4r)^2$ and
$b^2 = (1 + r_g/4r)^4$. However the explicit form of the metric functions
$a$ and $b$ is not essential. A coordinate transformation to another reference
frame moving at infinity with respect to the original one with velocity $V$
has been considered in ref. \cite{tpm}. The choice of coordinates made in this
book is not convenient for our purposes so we will take another coordinate
system which essentially coincides with that of ref. \cite{tpm} at 
asymptotically large distances from the gravitating center~\footnote{The 
inconvenience of the coordinate choice made in ref. \cite{tpm} is related to
the fact that for large $V$ the coordinate frame has a physical singularity
due to peculiarity of its motion near the source of gravity.}.

As one of the spatial coordinate lines we will choose the trajectory of 
the superluminal photon in this metric and the coordinate running
along this trajectory we denote as $l$. The other two, denoted $x_\bot$, 
are assumed to  be orthogonal (in three dimensional sense). 
We will consider only motion along this trajectory so that assume that
$x_\bot =0$. The metric along this trajectory can be written in the form:
\be 
ds^2 = A^2 (l) dt^2 - B^2 (l) dl^2  
\label{ds22}
\ee
Now let us go to a different coordinate frame which moves with respect to the
original one with velocity $V$ along $l$ at large distances from the center.
The corresponding coordinate transformation can be chosen as 
\be
t' = \gamma \left[ t -Vl - V f(l) \right]
\nonumber
\ee  
\be
l' = \gamma \left[l + f(l) -Vt  \right]
\label{t'l'} 
\ee 
where the function $f(l)$ is chosen in such a way so that the crossed terms
$dt'dl'$ do not appear in the metric. One can easily check that this can be
achieved if
\be
f(l) = \int^l dl \left( {B\over A} -1 \right)
\label{fl}
\ee 
In this moving frame the metric takes the form:
\be
ds^2 = A^2 \left( dt'^2 - dl'^2 \right)
\label{ds24}
\ee
where $A$ should be substituted as a function of $l'$ and $t'$.

The motion of the tachyonic photon in the original frame satisfies the 
condition 
\be
u= {Bdl \over Adt} = 1 + \delta u
\label{u}
\ee
The quantum correction to the speed of light $\delta u$ is assumed to 
be positive. As was shown in ref. \cite{dh} it is indeed the case for certain
photon polarization. In the case that the polarization turns out to 
correspond to subluminal propagation we can always put (in our gedanken
experiment) a depolarizor to change the polarization to the
superluminal one. Radially moving photons in Schwarzschild background has
normal velocity, $u=c$, so that we have to choose a trajectory with a nonzero
impact parameter. However in dilaton gravity, as shown in ref. \cite{cho}, 
a superluminal photon velocity is possible even for radial trajectories. In
this case the arguments proving an existence of time machine would be slightly
simpler.
 
The correction to the velocity of light is extremely small,
roughly it is 
\be
\delta u = { \alpha C_v r_g \rho\over r^4 m^2_e }  
\label{deltau}
\ee
where $r$ is the distance to the gravitating source and $\rho$ is the impact
parameter.
The coefficient $C_v$ is numerically small, 
about $\alpha /30\pi \approx 10^{-4}$, but the
main suppression comes form the enormously small factor $1/(rm_e)^2$. 

Correspondingly the time interval between emission and 
absorption of the superluminal photon is equal to
\be
t_2 - t_1 = \int_{l_1}^{l_2} dl \left( 1 - \delta u \right) (B/A)
\label{t2grav}
\ee
As follows from expression (\ref{t'l'}) this time interval in the primed system 
is  
\be
t'_2 -t'_1 = \gamma \left[ (1-V) \int_{l_1}^{l_2} dl  ( B / A)
  - \int_{l_1}^{l_2} dl \delta u (B/A) \right]
\label{t'2grav}
\ee
Clearly for sufficiently small $(1-V)$ this difference can be negative and
the signal can propagate into the past. Its
behavior as a function of the distance $(l_2-l_1)$ is similar to that found
above for the case of propagation between conduction plates (see 
eq.~(\ref{t2plate}).  

To arrange traveling backward in time we will do now the same trick as was
described above for the case of flat space-time. We assume that in the primed
system the photon detector moves symmetrically with respect to the photon source
i.e. it moves together with the attached to it another gravitating body
along the parallel path
with the same velocity but in the opposite direction. 
The arguments proving the possibility
of traveling backward in time in this frame exactly repeat the previous ones
for the propagation across the plates. 

It is worth noting that the gravitational field of the body moving together
with the detector influences the motion of the tachyon and its source 
and vice versa.
However for the motion with ultrarelativistic velocities this gravitational
field is concentrated in a very thin plane moving together with 
the body and perpendicular to the direction of the motion. Thus the influence
of this field on the magnitude of the tachyon velocity would be extremely short
and hence negligible. The direction of the tachyon motion might be changed quite 
significantly but this change could be immediately corrected by a mirror
or by an emission of a new tachyon with the same properties from the place
of the collision with the field.

\section{Conclusion}
 
Thus we have shown that quantum corrections to the photon 
propagation in a curved
space-time or across conducting plates, which lead to superluminal velocity,
would indeed permit travel backward in time, as one would naively expect.
This statement is in contradiction with the previously published 
papers~\cite{dh,gms} where it was claimed that though it would be possible
in a certain coordinate frame to arrive to detector earlier than the photon
had been emitted, still a return to the original emitter prior to the birth
of the photon was impossible. This statement was based on the assumed absence
of the Poincare invariance in the system under consideration. However, as one 
can see, the Poincare invariance still exists but in a slightly more complicated
form, namely one should consider the photon source or detector together with
the attached to each of them gravitating center as a single entity. After
this observation it becomes practically evident that traveling 
backward in time due to quantum corrections is indeed possible.

Thus we face the following dilemma, either time machine is possible in principle
or something is wrong in the conclusion of superluminal propagation due to 
quantum corrections (vacuum polarization). It was argued \cite{idn1,idn2} that
one can still has a consistent physics even if time travel is possible.
The other option that some unaccounted for effect may kill superluminal
propagation and permit to return to normality, looks more conservative and
for many people more natural. However at the moment it is not clear how this
cure can be achieved. The analysis of higher order corrections seems to 
support the present conclusion of superluminal propagation.

{\bf Acknowledgment.} We thank 
I. Khriplovich, E. Kotok, M. Marinov, H. Rubinstein, and
A. Vainshtein for discussion and criticism.
This work was supported in part by Danmarks Grundforskningsfond through its
funding of the Theoretical Astrophysical Center (TAC).

\newpage

{\large \bf Figure Captions:}
\vskip1cm
\noindent
{\bf Fig. 1.} $~~~$
Motion of tachyons in $(x-y)$-plane. A and B are tachyonic sources moving in
in the opposite directions as indicated. Tachyon emitted by A at the moment
$t_0$ moves backward in time and at the moment $t_1<t_0$ it is reflected
to B. The moment of collision of this tachyon with B triggers emission by
B of another tachyon which moves along $x$ again backward in time. At the
moment $t_3$ it is reflected back to A and hits it at the moment $t_4$ before 
it was emitted by A, $t_4< t_0$.
\newline  
{\bf Fig. 2.} $~~~$
Space-time trajectory of the first tachyon described in fig. 1.
\newline 
{\bf Fig. 3.} $~~~$
Space-time trajectory of the second tachyon described in fig. 1.
\newline 
{\bf Fig. 4.} $~~~$
Space-time trajectory of the tachyon moving between conducting plates. The
plates are assumed to be displaced along $y$ to avoid collision.

\newpage
 
\psfig{file=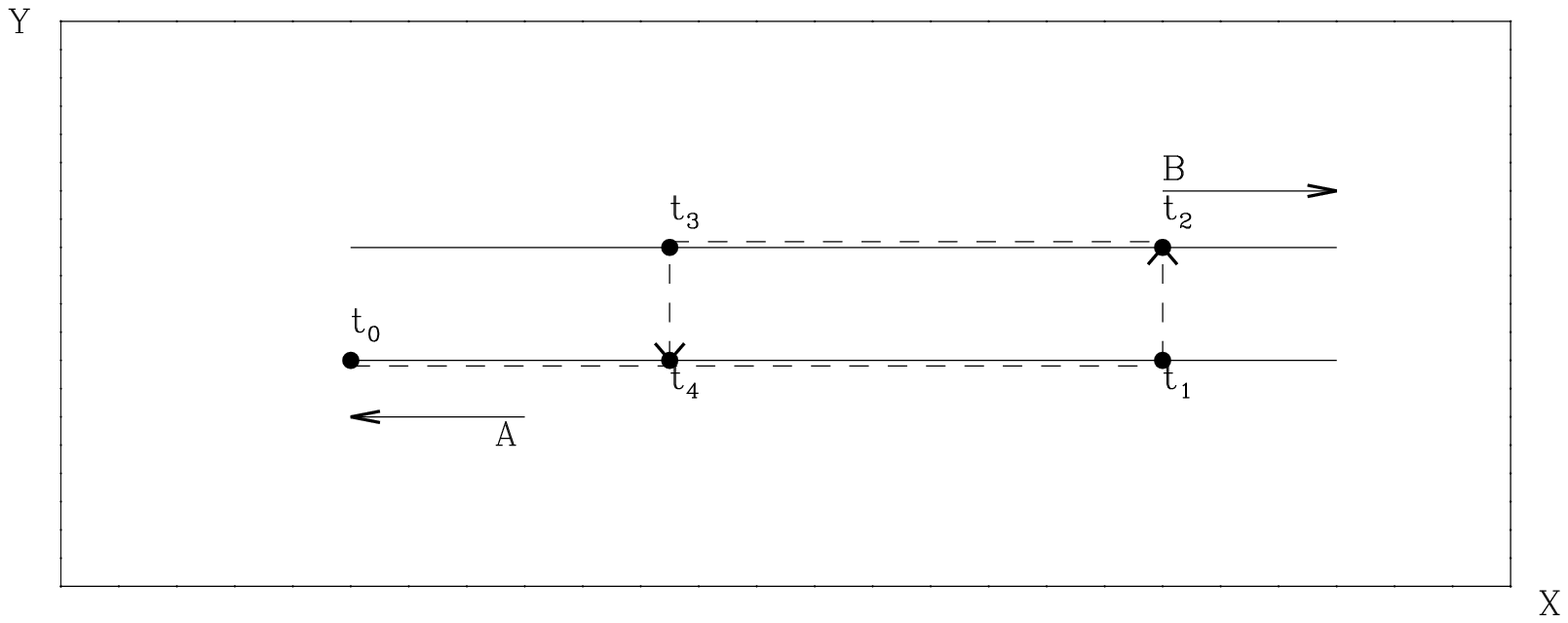,width=6in,height=4in}
\begin{center}
{\bf Figure 1.}
\end{center}

\newpage
 
\psfig{file=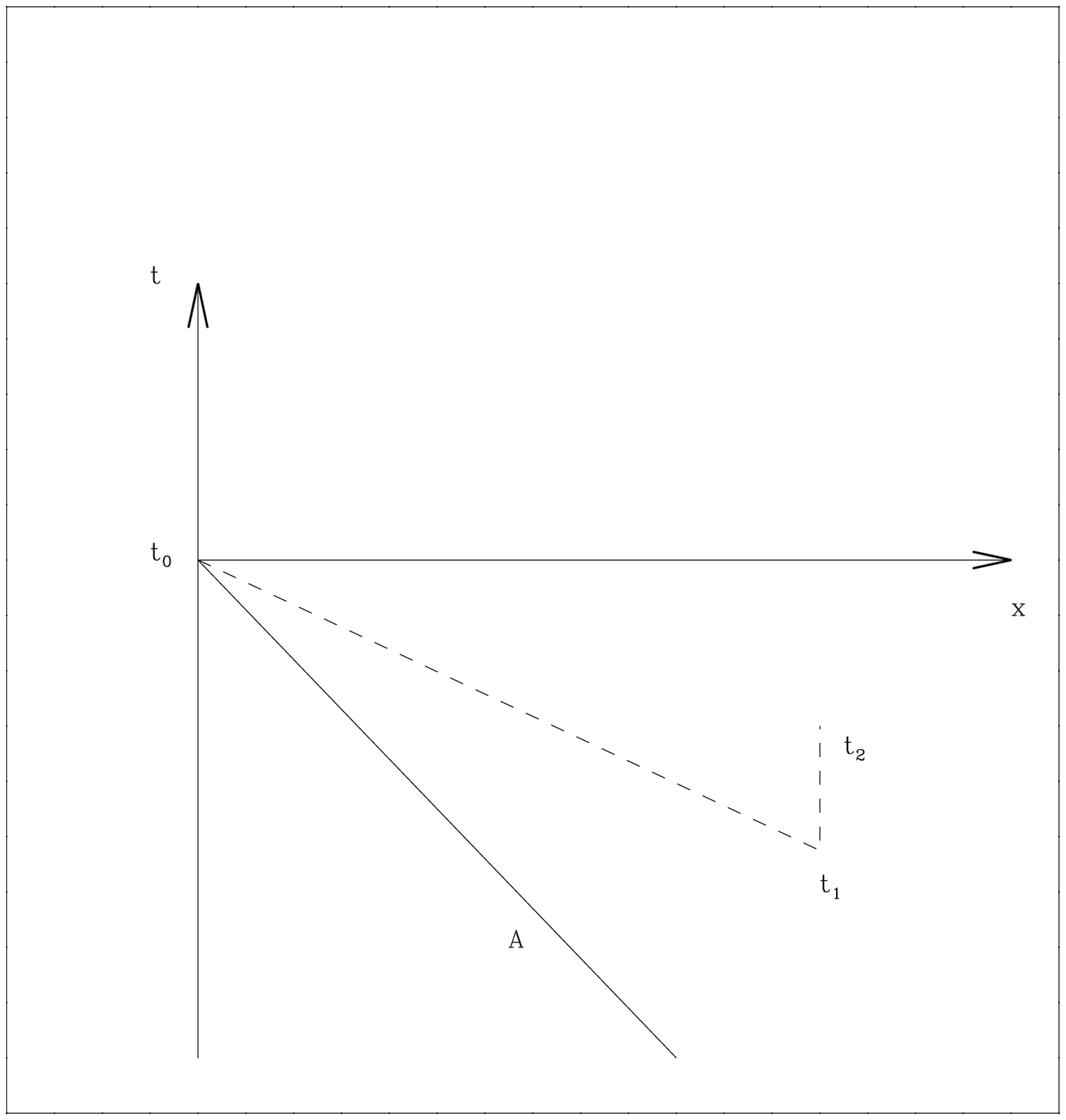,width=5in,height=7in}
\begin{center}
{\bf Figure 2.}
\end{center}
\newpage
 
\psfig{file=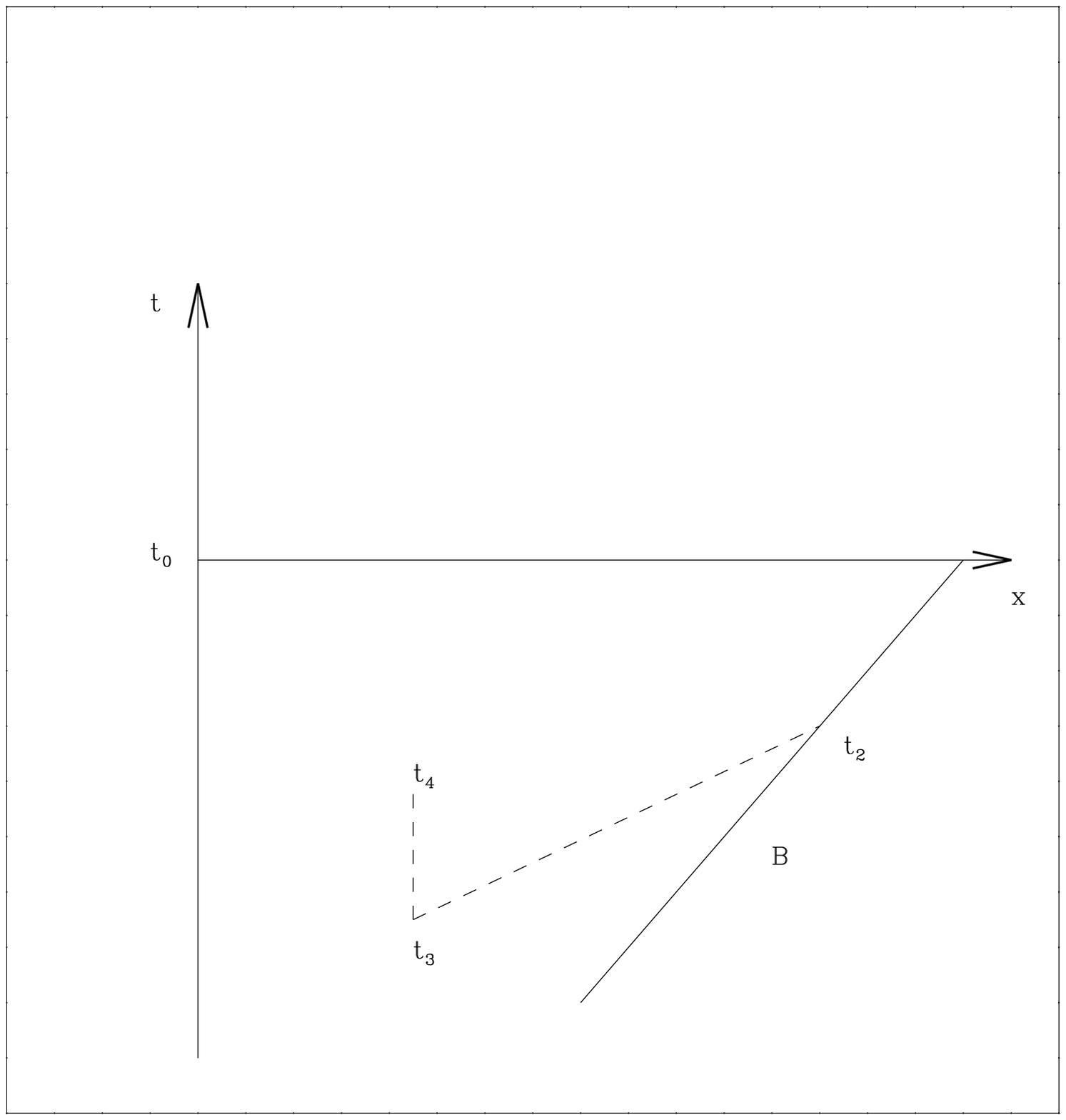,width=5in,height=7in}
\begin{center}
{\bf Figure 3.}
\end{center}
\newpage
 
\psfig{file=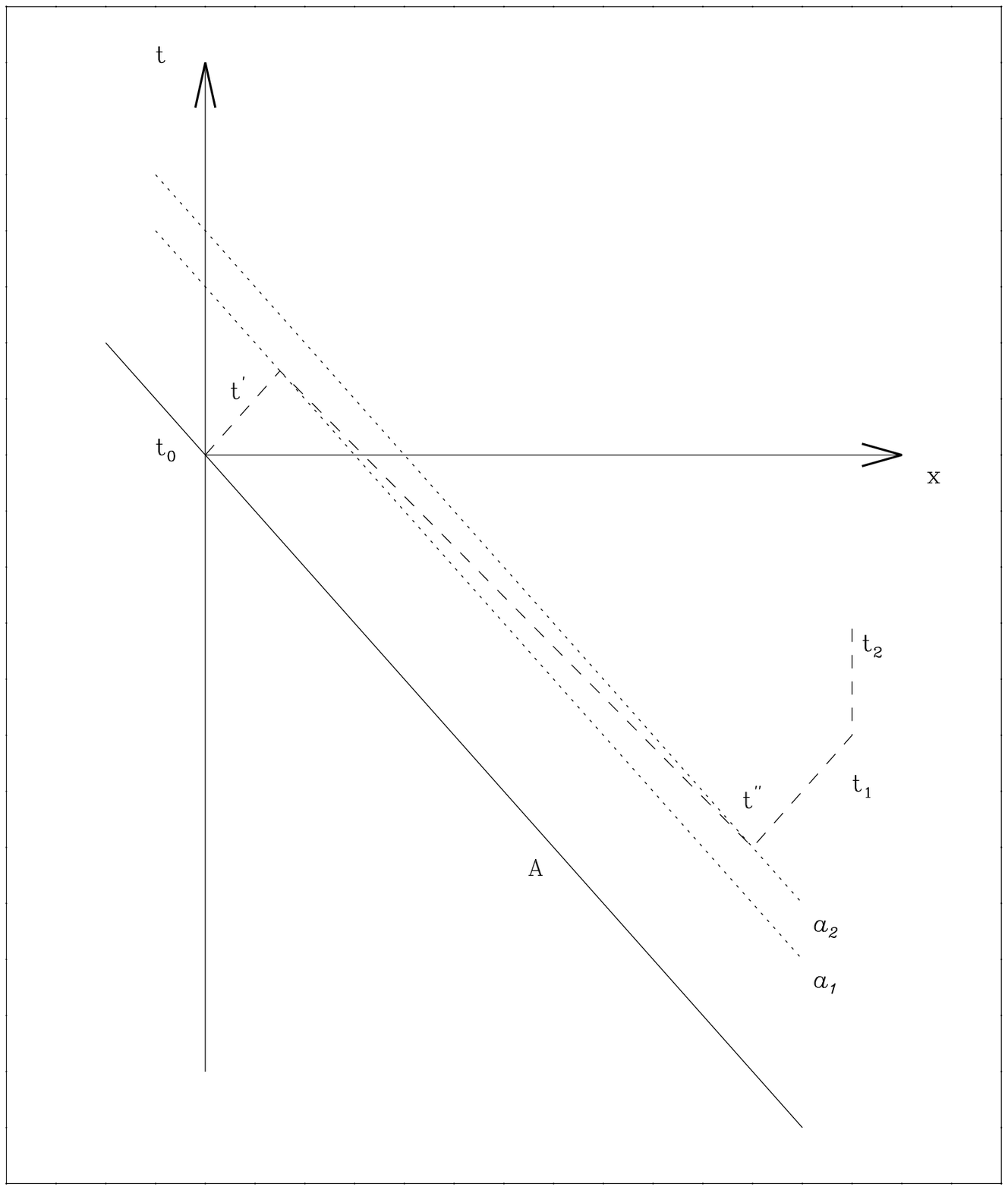,width=5in,height=7in}
\begin{center}
{\bf Figure 4.}
\end{center}
 
\end{document}